\documentclass[journal,twoside,web]{ieeecolor}
\usepackage{generic}
\usepackage{cite}
\usepackage{amsmath,amssymb,amsfonts}
\usepackage{algorithmic}
\usepackage{graphicx}
\usepackage{textcomp}
\usepackage{booktabs}
\usepackage{multirow}
\usepackage{color, hyperref}
\usepackage{comment}
\usepackage{enumerate}
\usepackage{hyperref}
\usepackage[dvipsnames]{xcolor}
\usepackage{pdfpages}
\usepackage{array}
\usepackage{stfloats}
\usepackage{url}
\usepackage{bm}

\usepackage{tabularx,booktabs}
\usepackage[T1]{fontenc}
\usepackage{subfigure}

\def\BibTeX{{\rm B\kern-.05em{\sc i\kern-.025em b}\kern-.08em
    T\kern-.1667em\lower.7ex\hbox{E}\kern-.125emX}}
\begin{document}
\title{Construction of an Organ Shape Atlas Using a Hierarchical Mesh Variational Autoencoder}
\author{Zijie Wang, Ryuichi Umehara, Mitsuhiro Nakamura, Megumi Nakao \IEEEmembership{Member, IEEE}
\thanks{This study involved human subjects. Approval of all ethical and experimental procedures and protocols was granted by Kyoto University Certified Review Board (approval number: R1446), and they were performed in line with the Declaration of Helsinki. This study was supported by JSPS Grant-in-Aid for Scientific Research (A) 24H00795 and (B) 22H03021.}
\thanks{Z. Wang and R. Umehara are with the Graduate School of Informatics, Kyoto University. Yoshida-Honmachi, Sakyo, Kyoto, 606-8501, JAPAN.}
\thanks{M. Nakamura and M. Nakao are with the Graduate School of Medicine, Kyoto University. Kawahara-chou 53, Shogoin, Sakyo, Kyoto, 606-8507, JAPAN.}}
\maketitle

\begin{abstract}
An organ shape atlas, which represents the shape and position of the organs and skeleton of a living body using a small number of parameters, is expected to have a wide range of clinical applications, including intraoperative guidance and radiotherapy. Because the shape and position of soft organs vary greatly among patients, it is difficult for linear models to reconstruct shapes that have large local variations. Because it is difficult for conventional nonlinear models to control and interpret the organ shapes obtained, deep learning has been attracting attention in three-dimensional shape representation. 

In this study, we propose an organ shape atlas based on a mesh variational autoencoder (MeshVAE) with hierarchical latent variables. To represent the complex shapes of biological organs and nonlinear shape differences between individuals, the proposed method maintains the performance of organ shape reconstruction by hierarchizing latent variables and enables shape representation using lower-dimensional latent variables. Additionally, templates that define vertex correspondence between different resolutions enable hierarchical representation in mesh data and control the global and local features of the organ shape. 

We trained the model using liver and stomach organ meshes obtained from 124 cases and confirmed that the model reconstructed the position and shape with an average distance between vertices of 1.5 mm and mean distance of 0.7 mm for the liver shape, and an average distance between vertices of 1.4 mm and mean distance of 0.8 mm for the stomach shape on test data from 19 of cases. We also performed two sets of shape interpolation between two cases for the liver and stomach shapes, and compared the shape interpolation between the proposed method and conventional methods using principal component analysis (PCA). In the shape interpolation representation, the proposed method continuously represented interpolated shapes, and by changing latent variables at different hierarchical levels, the proposed method hierarchically separated shape features compared with PCA.
\end{abstract}


\section{Introduction}
\IEEEPARstart{T}{hree-dimensional(3D)} imaging is widely used in clinical medicine, including surgery and radiotherapy, to obtain morphological structures of patient-specific organs. The anatomical structures of organs, including vasculature and tumors, are represented in three dimensions by extracting the region of interest from computed tomography (CT) or magnetic resonance images before treatment. However, because of the physical limitations of equipment and measurement time, it is difficult to obtain the exact shape of an individual patient's organs at the time of treatment using measurement alone. Various studies have been conducted to estimate the 3D shape only from images that can be captured in real time\cite{Hvid18}\cite{Posiewnik19}\cite{Magallon19}.

For example, in endoscopic surgery, two-dimensional (2D) images of target organs taken by an endoscopic camera are generally used to provide information on the intraoperative state of the organs. However, it is difficult to grasp the anatomical structure of the organs because of the high uncertainty caused by organ deformation and the limited field of view during surgery\cite{Tokuno20}. For this reason, research is being conducted to align the 3D shapes of organs that were imaged prior to treatment with the 2D images of the organs to be operated on. To achieve 3D shape reconstruction and alignment based on such low-dimensional and incomplete information, prior knowledge of higher-dimensional organ shapes is required data-driven representations of organ shapes, called organ shape atlases, have been explored. An organ shape atlas is a model that represents geometric information, such as the position and shape of organs in a living body and their inter-individual variability, using a small number of parameters, and is widely used in medical image analysis and biomedical engineering fields\cite{Soliman17}.

Organ shape atlases have been formulated based on the concept of the statistical shape model (SSM) \cite{Heimann09} and can be broadly classified into image-based methods \cite{Oh17}\cite{Sotiras13} and mesh-based methods \cite{Nakao21}. Image-based methods directly use 3D medical images and construct an SSM based on the spatial distribution of the existence probabilities of target organs. Because the pixel is a grid-like data structure, it has the advantage of being easy to process on a computer, for example, feature calculation and filtering. However, it is difficult to represent complex organ shapes at high resolution from the viewpoint of computation time because the amount of data in a 3D image increases in proportion to the cube of the image size. additionally, it is not easy to process nonlinear and discontinuous shape changes and deformations, such as rotation and torsion between organs, using only a pixel-based shape description. The mesh-based method requires preprocessing to create a mesh of organ shapes from 3D images, but it can represent complex shapes using a very small amount of data consisting of vertices and edges, which is characterized by its easy handling of deformation. However, because a mesh is a graph structure and not spatially aligned data such as pixels, it is difficult for it to describe processes such as feature calculation, sampling, and reconstruction. 

In this study, we use the mesh-based model, and focus on its shape representation and generalization performance. In our previous study, we attempted to construct an organ shape atlas using a deep learning model that combines a variational autoencoder (VAE) \cite{Kingma13} and graph convolutional network (GCN) \cite{Kipf16}\cite{Umehara23}. However, the GCN requires the same number of latent variables as the input mesh, and it is difficult to analyze the latent variables and control the generated organ shapes. In this study, we propose an organ shape atlas based on mesh VAE (MeshVAE) with hierarchical latent variables. The proposed method (Proposed) adds two new technical elements to shape representation using MeshVAE \cite{Umehara23} to represent the complex shapes of biological organs and nonlinear shape differences between individuals. The first is the hierarchization of latent variables. By hierarchizing latent variables, we expect to achieve the overall shape representation with lower-dimensional latent variables while maintaining the performance of organ shape reconstruction. The second technical element is the construction of organ shape templates with vertex correspondence between different resolutions using mesh subdivision. Mesh subdivision enables a hierarchical representation of mesh data, and the control of global and local features of organ geometry. By incorporating these two elements into MeshVAE, we aim to achieve an organ shape atlas with both high reconstruction performance and interpretability.

To validate the effectiveness of these technological elements, we performed model training using liver and stomach mesh groups generated from 3D CT images of 124 patients, and compared the difference in reconstruction performance with that of several conventional methods. we extracted the latent variables at different hierarchical levels from the two organ meshes. We reconstructed the organ shapes from the new latent variables calculated by the linear interpolation of the latent variables and compared the interpolation results of the organ shapes expressed by each hierarchy.

\section{Related Research}
\subsection{Data Structure of the Organ Shape Atlas}
In a wide range of fields, such as computer vision, graphics, and shape analysis, three types of data structures, that is,-3D image, point cloud, and mesh are generally used to represent 3D shapes on computers and also used in the shape representation of organ shape atlases. A 3D image is composed by superimposing 2D sliced images in the 3D direction. It represents an object using the spatial distribution of scalar values or vectors, with voxels as the smallest element, and has been used in organ atlases, such as the 3D atlas of the brain in brain function analysis\cite{Wu16}\cite{Choy16}\cite{Jimenez16}. The data obtained using 3D imaging can be directly used, and is the most frequently used format in medical image analysis. By contrast, the data volume of a 3D image increases in proportion to the cube of the resolution. Additionally, artifacts such as partial volume effects may occur when converting to low-resolution 3D images. Therefore, there is a limit to the representation of high-resolution geometry in voxel format, particularly in scenarios in which real-time prediction is required.

Point cloud data represent a set of points sampled from a 3D region. The voxel can be regarded as a generalized data format in a 3D image. Various methods for reconstructing 3D point cloud information from only one image have been studied in recent years \cite{Valsesia18}\cite{Achlioptas18}\cite{Yang19}. However, some organs, such as the brain and intestines, are self-adjacent, and many organs have fine irregularities and branches, such as vertebrae and vascular vessels. Point cloud data are not suitable for accurately describing complex shapes and deformations because they do not have explicit relationships between vertices or surface information about shapes.

Mesh data are a data format that represents a 3D surface using a set of points and their connecting edges. Mesh data represent a type of undirected graph and are widely used in computer graphics, shape analysis, and human body shape modeling. Complex shapes can be represented using a very small amount of mesh data consisting of vertices and edges, and deformation is easy to handle. However, mesh data require preprocessing for conversion from 3D images, and it is difficult for such data to describe processes such as feature calculation and reconstruction. In this study, we focus on mesh-based modeling methods among organ shape atlas modeling methods.

\subsection{Organ Shape Atlas Modeling Methods}
Next, we outline the modeling method for the organ shape atlas. Computational models used for organ shape atlases can be broadly classified into two categories: linear models and nonlinear models.
\subsubsection{Linear models}
In a mesh-based SSM, features are extracted from a vertex-correlated mesh of organ shapes and added to the average shape by summing the weights to represent arbitrary organ shapes. Principal component analysis (PCA)\cite{Nakao19}\cite{Cootes95}\cite{Nakamura} is commonly used for linear SSM. In the method using PCA, principal components that represent shape variations are calculated from the mesh vertices in the dataset, and the average shape is added to the sum of their weights to represent the organ shape.

Principal components with larger eigenvalues can express the shape differences inherent in the data to a greater extent; that is, the larger the eigenvalues of principal components, the larger the change in the mesh vertex group expressed by changing the value of the coefficient. Because all principal components are calculated from all the vertices on the mesh, changing one principal component in PCA changes the entire represented organ geometry. This feature makes interpretation difficult when considering applications such as shape analysis and diagnosis. To overcome this challenge, research has been conducted using independent component analysis (ICA)\cite{Hyvarinen01} and sparse PCA\cite{Sjostrand07}. Üzümcü et al.\cite{Uzumcu03} constructed an SSM for left ventricular geometry using ICA and achieved local shape changes in several test cases that could not be represented using PCA, but did not achieve the same dimensionality reduction as PCA. Sjostrand\cite{Sjostrand07} et al. constructed an SSM for the shape of the corpus callosum using sparse PCA. By reducing the 3D shape to a lower dimension, they analyzed the relationship between clinical data, such as age and gender, and local features of the shape. Additionally, they analyzed which part of the shape of the corpus callosum is related to changes in the values of clinical data. Thus, the linear method can represent the 3D shape in an interpretable form with few parameters. By contrast, the linear method has difficulty in representing locally large deformations and nonlinear deformations, such as torsion and rotation.

\subsubsection{Nonlinear models}

Studies have been conducted on an SSM using nonlinear methods to solve the problems of linear methods. Sozou et al. constructed an SSM of chromosomes using nonlinear methods, such as polynomial regression\cite{Sozou94} and a multilayer perceptron\cite{Sozou95}, and succeeded in capturing nonlinear deformations that cannot be represented by linear methods for some shapes. Other methods using nonlinear models, such as kernel PCA\cite{Nakao21}\cite{Tayler11}, have also been proposed. In recent years, deep learning, which is one of the frameworks of nonlinear methods, has attracted attention regarding 3D shape representation because of its good representation performance. Tan et al.\cite{Tan18} created a VAE model (MeshVAE) that consisted of fully connected layers for the purpose of creating a generative model that can analyze and synthesize mesh geometry. The VAE model can handle mesh data by calculating and inputting rotation-invariant mesh difference features\cite{Gao16} from the mesh data. However, the model with fully connected layers has a large number of parameters and its generalization performance is poor.

To solve the above issues, research has been conducted using GCN as a derivative model of MeshVAE to improve the shape representation capability while reducing the number of model parameters\cite{Ranjan18}\cite{Yuan20}\cite{Sara22}. Ranjan\cite{Ranjan18} et al. proposed a convolutional mesh autoencoder (CoMA) for the shape representation of a face shape mesh, which combines GCN and a pooling method for sampling vertices, and confirmed that it can represent complex face shapes with low-dimensional latent variables. However, the sampling-based pooling method does not aggregate all local neighborhood information when reducing the number of vertices and cannot uniformly aggregate the information on the vertices. Yuan\cite{Yuan20} et al. proposed a new pooling method that removes edges and merges their connected vertices to construct a MeshVAE with high generalization performance. In this pooling method, the latent variables obtained may have different meanings among the data because the method of edge deletion varies depending on the geometry.

Sara\cite{Sara22} et al. constructed MeshVAE, which introduces convolution and pooling methods for mesh faces, thereby aiming at a more general framework that can learn transitions for mesh shapes with different connection relations. In the framework proposed by Sara et al., the input mesh is simplified and converted into a semi-regular mesh shape using preprocessing loop subdivision. Within a small region, the number of adjacent vertices for each vertex is always 6, except at the edges, which successfully defines a convolution operation for small regions similar to that for images. By applying the same preprocessing to other meshes with different topology, it is possible to define convolution operations using learned weights. However, semi-regular meshes gather a large number of vertices at most vertices, which makes it difficult to represent parts of the shape that have large curvature. Additionally, mesh transformation may cause errors with the original shape.

Recently, studies have been conducted to use VAE as an organ shape atlas. We aimed to construct an organ shape atlas using MeshVAE\cite{Umehara23}. However, the use of GCNs makes the dimensions of input-output and latent variables the same, which does not contribute to reducing the number of dimensions of latent variables and makes the analysis difficult. Beetz\cite{Beetz22} et al. drew attention to the versatility of MeshVAE and constructed an organ shape atlas using the 3D mesh data of the left ventricle of the heart using CoMA by Ranjan\cite{Ranjan18} et al. to demonstrate its applicability in medicine. This suggests that it can be applied to the generation of an unknown left ventricular shape and the prediction of medical examination prognosis from dimensionally compressed information as a latent variable. Nonlinear methods can now represent nonlinear deformations, such as rotation and torsion, which are difficult to represent using linear methods; however, they are prone to overfitting if there is a limited amount of data.

\subsection{Contributions of this study}
The aim of this study is to construct an organ shape atlas with an improved ability to represent the complex shapes of organs by introducing hierarchical latent variables into MeshVAE. We expect the VAE with hierarchical latent variables to achieve both the representational performance and interpretability of 3D shapes, which has been a problem with nonlinear methods, while acquiring nonlinear shape representations, such as torsion and rotation, which have been difficult to achieve with linear models. To the best of our knowledge, there are no known applications of MeshVAE to hierarchical models and their organ shape atlas.

The contributions of this study are as follows:
\begin{itemize}
    \item use of multi-level template meshes in an organ shape atlas;
    \item proposed use of MeshPooling to convert between high- and low-resolution meshes; and
    \item proposed use of a MeshVAE model with a hierarchical structure for latent variables.
\end{itemize}

\section{Method}
\subsection{Overview of the proposed method}
In this study, we aim to develop a framework for MeshVAEs that represents 3D shapes hierarchically using hierarchizing latent variables. Our assumption is that the organ meshes that are the input data for the proposed model and are to be reconstructed have the same number of vertices and edges, and we assume that they represent a set of well-aligned meshes with vertex correspondence. However, in general, when organ regions are extracted from patient-specific 3D images and organ shapes are represented, for example, by surface meshes,  the number of vertices and edges varies depending on the conditions at the time of generation. Therefore, preprocessing is performed to align the number of vertices and edges, and their connection relationships between meshes in the dataset.

\begin{figure}[t]
	\flushright
	\includegraphics[width=1.0\linewidth]{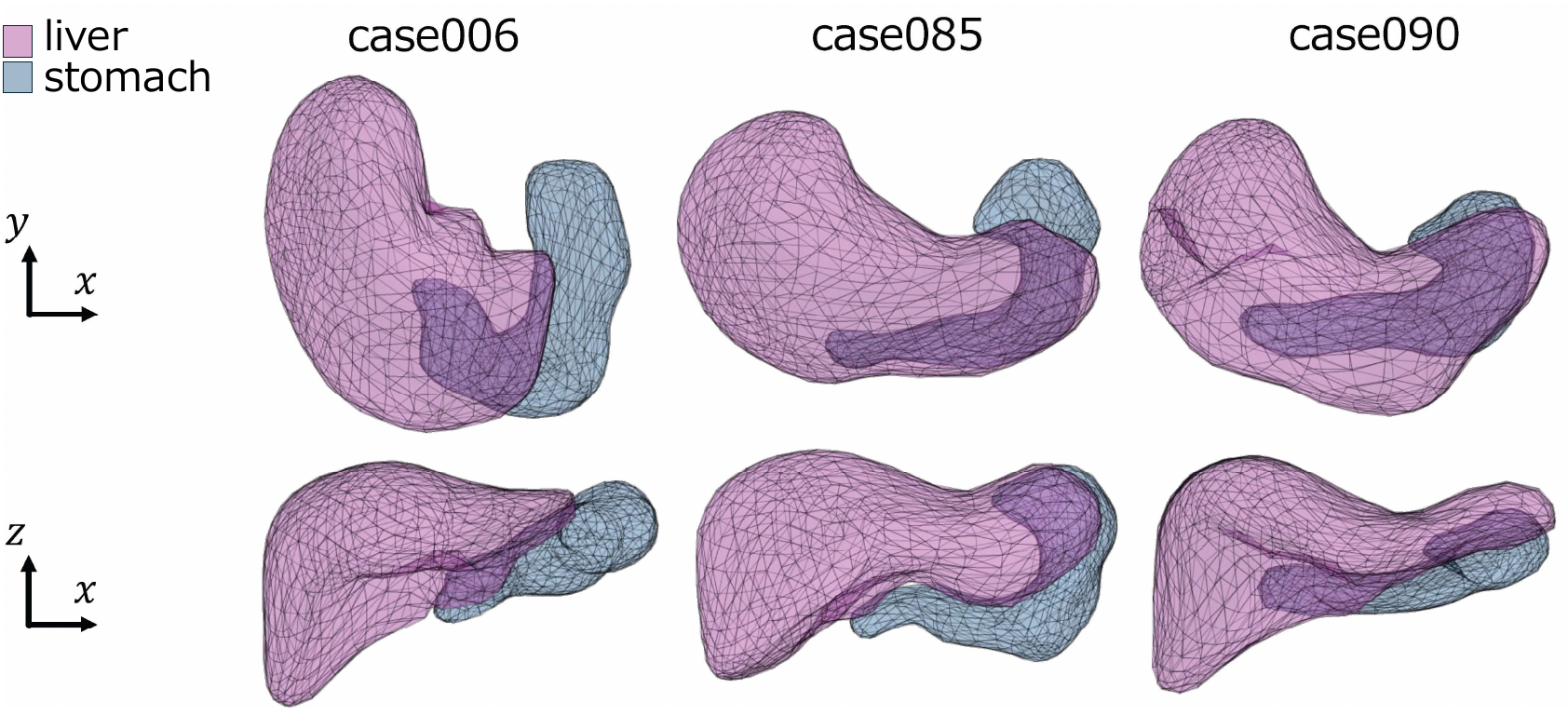}
	\caption{Three examples of the liver and stomach mesh after positioning in this study.}
	\label{sample}
\end{figure}

A group of meshes with the same number of vertices and the same phase structure can be prepared by applying deformable mesh registration (DMR)\cite{Nakao21} as a preprocessing step to the organ meshes of all cases. In DMR, the organ shape of a case with an average shape is first selected as a template from the dataset. The template is then deformed and aligned so that its shape matches the mesh in the remainder of the dataset as closely as possible. In\cite{Nakao21}, the vertex positions were gradually changed to constitute the surface of the target mesh while maintaining the initial topology of the mesh. This produced a mesh that approximated the organ shape of all cases while having the same number of vertices and vertex connectivity as the template. In the pre- and post-aligned meshes, the organ geometry of each target case could be represented using an average Hausdorff distance (HD) of 1.1 mm and Dice coefficient of 98.1\% for the liver, and with an average HD of 0.9 mm and Dice coefficient of 97.0\% for the stomach. The post-aligned mesh data used in this study was found to represent a 3D shape with sufficient accuracy. Three examples of the actual aligned mesh data used in this study are shown in Figure \ref{sample}, where the pink mesh is the liver and the blue mesh is the stomach.

\begin{figure*}[htbp]
	\centering
    \includegraphics[width=1.0\linewidth]{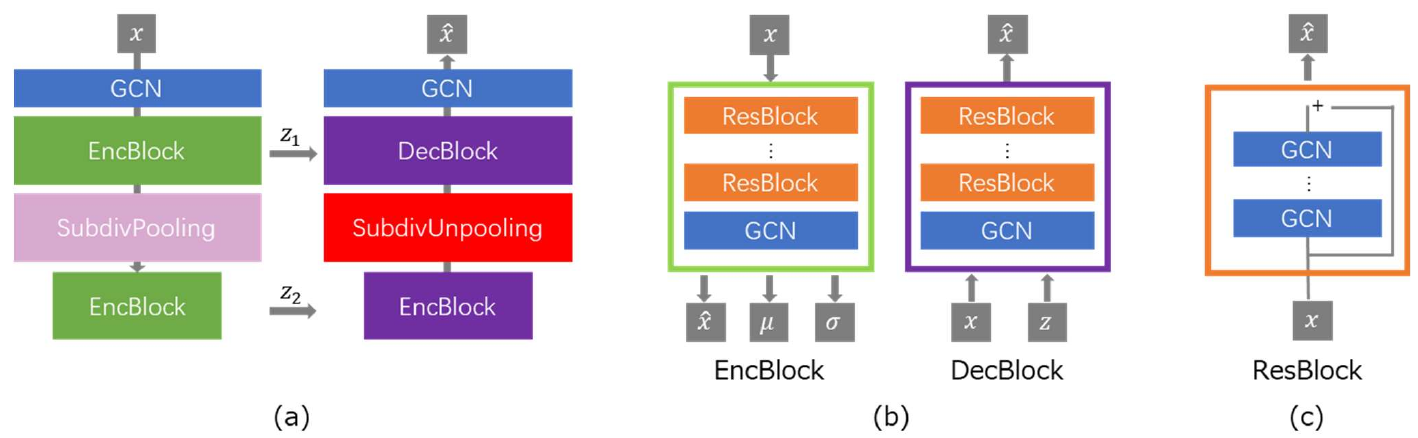}
	\caption{Proposed model: (a) framework overview, (b) structure of EnBlock and DecBlock in the framework, (c) structure of ResBlock in EnBlock and DecBlock. The model further uses a pooling/unpooling layer to achieve hierarchical latent variables.}
	\label{fig1}
\end{figure*}

Figure \ref{fig1} shows the model of the designed hierarchical MeshVAE. The designed hierarchical MeshVAE consists of an encoder, which includes an encoder block (EncBlock) and subdivision neighbor pooling (SubdivPooling), and a decoder, which includes a decoder block (DecBlock) and subdivision neighbor unpooling (SubdivUnpooling). The EncBlock and DecBlock use the residual connection block (residual connection block, ResBlock) constructed in the GCN. The learning of the weights in this model is achieved by minimizing the loss function calculated from the inputs and outputs of the model.

The proposed model implements SubdivPooling/Unpooling corresponding to the graph structure of the mesh, which allows the simultaneous creation in the model of lower-dimensional and lower-resolution latent variables, in addition to high-resolution latent variables with the same dimensions as the input mesh. In convolutional neural networks  in the field of imaging, pooling/unpooling is used to reduce the dimensionality of features and to aggregate global information\cite{Voulodimos18}. In the pooling operation on an image, it is easy to map pixels before and after pooling because of the constraint that the number of neighboring pixels is always 8, except for pixels located at the edges of the image. By contrast, in meshes, the number of vertices adjacent to a vertex is different for each vertex; hence, it is not easy to determine the correspondence between the vertices before and after pooling for an arbitrarily generated mesh. To solve this problem, various studies on mesh pooling have been conducted, which can be broadly classified into the following three types.

\begin{itemize}
    \item vertex sampling method \cite{Ranjan18};
    \item deleting edges and merging two connected vertices \cite{Yuan20}; and
    \item merging surfaces \cite{Sara22}.
\end{itemize}

The vertex sampling method\cite{Ranjan18} involves sampling vertices against a template mesh. During sampling, the vertices to be removed are selected so that their shape changes as little as possible before and after preprocessing. In the simplified mesh, vertices can be inserted at the center of gravity of the three nearest unsampled vertices from the sampled vertex, thus making the mesh vertex correspond between different hierarchical levels.

In the method of deleting an edge and merging two connected vertices\cite{Yuan20}, the edge is deleted so that its shape changes as little as possible before and after processing, and the two vertices connected to the edge are merged as one vertex for pooling. Unpooling allows the correspondence of potential vertices before and after pooling by inserting edges between the vertices merged using pooling.

In the above two methods, the pooling operation is highly dependent on the shape of the template mesh. It is difficult to learn to aggregate global features when the shapes in the template and patient-specific mesh are very different, such as the abdominal organ shapes targeted in this study. In the method of merging faces\cite{Sara22}, pooling is performed by merging four faces, that is, one face and three adjacent faces, into one face, and unpooling is performed by dividing a triangle into four triangles by inserting a point on each side of the triangle. However, a limitation exists in the mesh to which this method can be applied for pooling to the entire mesh from Euler's polyhedral theorem. This is because the number of faces changes by a factor of four before and after pooling by merging the faces.

In this study, we propose SubdivPooling/Unpooling based on the method of merging faces. To solve the problems in the method of merging faces, we introduce a hierarchical template that defines the topology of the mesh before and after pooling. We also propose SubdivPooling/Unpooling for hierarchical templates, which realizes pooling by merging faces so that vertices are not localized using pooling or unpooling.

Next, we consider how to encode from the mesh to the latent variables and how to decode from the latent variables. As a deep generative model, VAE was originally a widely studied model in the field of imaging. In recent years, attempts have been made to hierarchize the latent variables to be encoded. Razavi et al.\cite{vq-vae2} confirmed an improvement in the expressive power of VAE models by making the latent variables hierarchical in vector quantized-VAE\cite{vq-vae}. Child et al.\cite{Child20} also suggested that by deepening the layers of the VAE and hierarchizing the latent variables, it is possible to separate the phenotypes to be learned for each layer.

In the present study, we newly introduce the concept of hierarchical latent variables, which has been explored in the field of VAE, into MeshVAE, which is a deep mesh generation model; that is, we achieve the hierarchization of MeshVAE by directly passing the high-resolution and low-resolution latent variables defined by the encoder to the decoder, which aims to separate and represent the organ geometry into global and local features. Details of the proposed hierarchical template and latent variable hierarchy are described in Sections B and C, respectively.

\subsection{Hierarchical template}

From Euler's polyhedral theorem, the relationship between the number of elements $f$ and the number of vertices $v$ of the triangular mesh used in this study is given by

\begin{equation}
    2v - f = 4.
    \label{eq: Euler}
\end{equation}

\begin{figure}[tbp]
	\centering
    \includegraphics[width=1.0\linewidth]{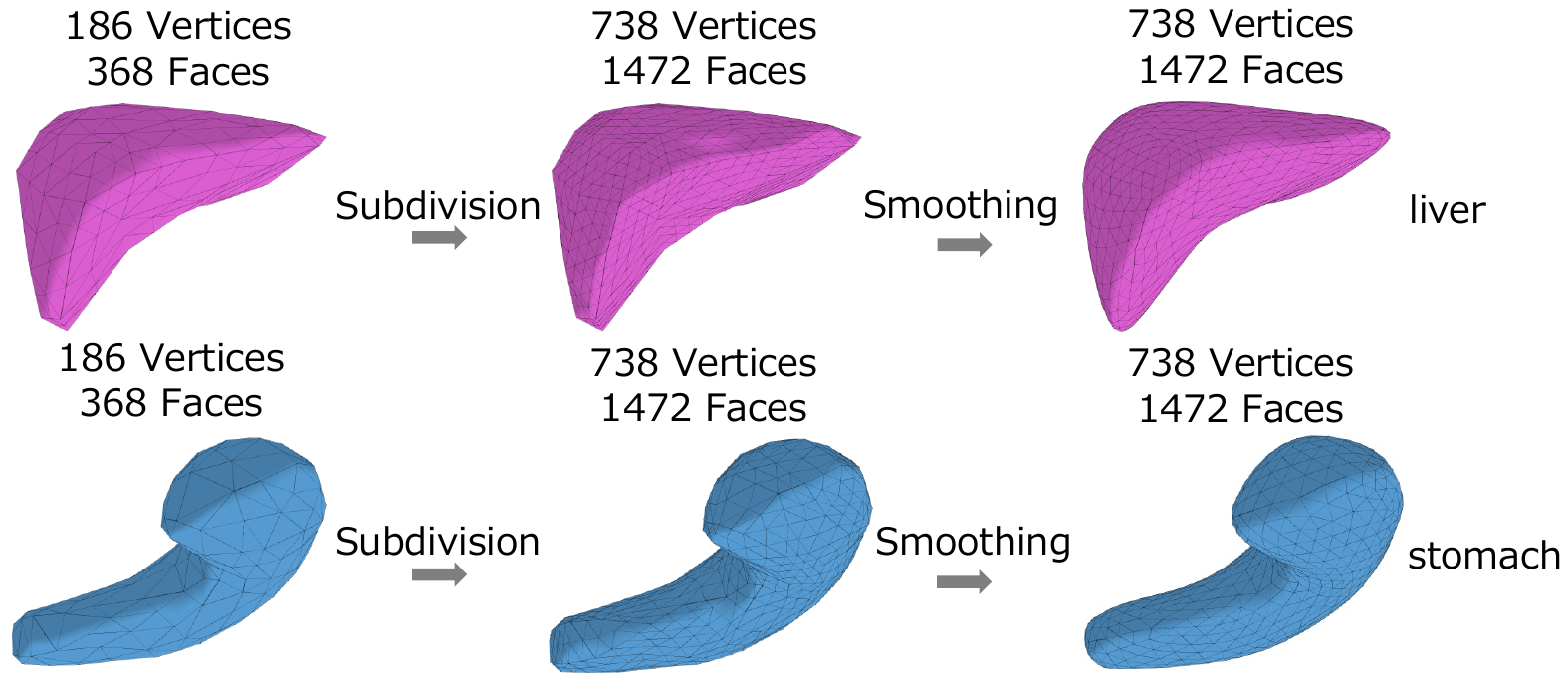}
	\caption{Hierarchical template creation procedure: generated using subdivision and smoothing from a template representing the mean shape.}
	\label{h-template}
\end{figure}

If SubdivPooling is applied to all elements, the number of elements is reduced by a factor of four; hence, SubdivPooling may not be applicable to all meshes. It is also not obvious which four faces are merged into one face in SubdivPooling. To solve these two issues, in this study, we define the correspondence of vertices before and after SubdivPooling using a hierarchical template created based on the loop subdivision method\cite{Stam98}. Figure \ref{h-template} shows how the hierarchical template is created.

First, a low-dimensional template consisting of a small number of vertices is obtained using mesh simplification from a template representing the mean shape\cite{Nakao21}. Loop subdivision is applied to the low-resolution template and smoothing is performed to obtain a high-resolution template with four times the number of triangle elements. In Figure \ref{h-template}, we started with a low-dimensional template consisting of 386 triangular elements with the mean shape of and subdivided to yield high-resolution templates with 1,472 elements. From the vertex correspondence between the low-resolution and high-resolution templates, it was possible to uniquely determine the vertices to be sampled in SubdivPooling/Unpooling.

As described in Section A, the original meshes obtained from patient-specific 3D CT images do not have the same number of vertices and connection relationships between vertices across patients. Therefore, using the high-resolution template as the initial template and performing mesh deformation registration\cite{Nakao21} on the organ meshes of all cases, a set of meshes with the same number of vertices and the same topological structure, but with patient-specific shapes, can be obtained. Vertex correspondences between the different hierarchies have already been obtained using the mesh subdivision described above, and the vertex group of the low-resolution mesh is a subset of the vertex group of the high-resolution mesh. Therefore, once the vertex positions are determined by high-resolution mesh registration, the vertex positions of the low-resolution mesh are also uniquely determined.

\subsection{Hierarchical MeshVAE}
\subsubsection{SubdivPooling/Unpooling}

The model hierarchizes the latent variables and separates the shape features to be learned for each hierarchy, which improves the reconstruction performance of the model and facilitates the analysis of the latent variables. To achieve this, in addition to using the high-resolution latent variables of the same dimension as the input-output mesh vertex groups, it is necessary to prepare lower-resolution latent variables of a lower dimension. Additionally, they need to be uniquely transformed in both directions. To satisfy these two requirements, in this study, we propose SubdivPooling/Unpooling using the hierarchical templates introduced in the previous section.

\begin{figure}[tpb]
	\centering
	\includegraphics[width=0.8\linewidth]{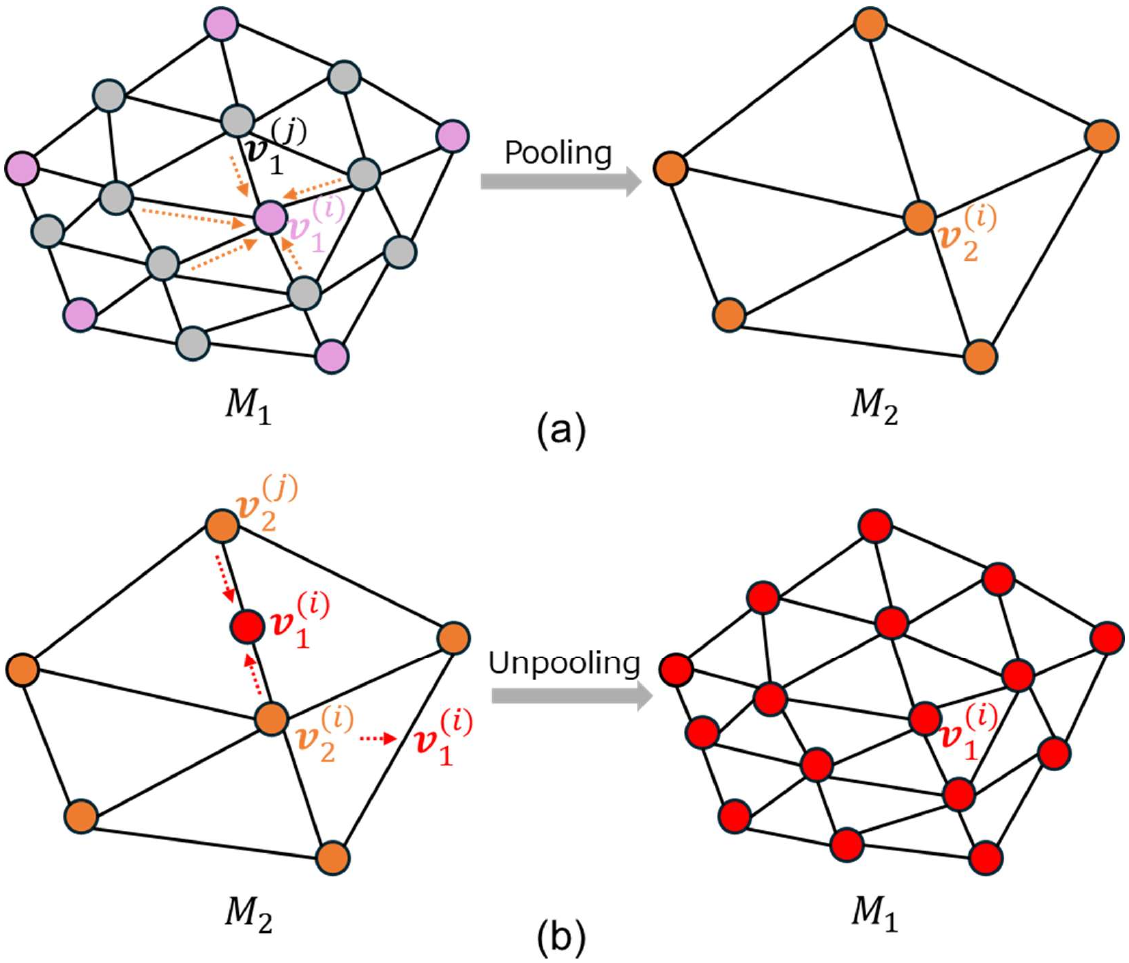}
	\caption{Conceptual diagram of: (a) subdivision neighbor pooling, (b) subdivision neighbor unpooling.}
	\label{pooling}
\end{figure}

Various aggregation functions can be used for the neighboring values in pooling; however, as the organ shape is represented by the mesh vertices, pooling/unpooling uses the average of the neighboring vertices as the aggregation function so that the shape is as consistent as possible before and after pooling. Specifically, as shown in Figure \ref{pooling}, for SubdivPooling, the information held by the output vertex (orange) is the average of itself (pink) and its neighboring vertices (gray) in the input, whereas for SubdivUnpooling, a new vertex is inserted at the mid-point of the mesh edge.

For high-resolution mesh $M_1$ and low-resolution mesh $M_2$ after pooling, the vertex sets are $V_1=\{\boldsymbol{v}_{1}^{(i)}\}$ and $V_2=\{\boldsymbol{v}_{2}^{(i)}\}(i=1, 2, \ldots)$, respectively. The transformations of the features in SubdivPooling and SubdivUnpooling are defined by the following, respectively:

\begin{equation}
    \boldsymbol{v}_2^{(i)} = Pooling(V_1, i) = \frac{1}{|\mathcal{N}_i| + 1}\sum_{j \in{\mathcal{N}_i}} \boldsymbol{v}_1^{(j)}
    \label{eq-pooling}
\end{equation}

\begin{equation}
    \boldsymbol{v}_1^{(i)}= Unpooling(V_2, i) = \left\{
    \begin{aligned}
        &\boldsymbol{v}_2^{(i)}, \text{if } \boldsymbol{v}_1^{(i)} \in{V_2}\\
        &\frac{1}{2}(\boldsymbol{v}_2^{(i)} + \boldsymbol{v}_2^{(j)}), \text{otherwise}
    \end{aligned}
    \right.
    \label{eq-unpooling}
\end{equation}

where $\mathcal{N}_i$ is the set of vertex numbers of the vertices adjacent to vertex number $i$. In \eqref{eq-unpooling}, $\boldsymbol{v}_2^{(j)}$ is the vertex number of the vertex which is adjacent to vertex $i$ in $V_2$.

\subsubsection{Hierarchical latent variables}
In this section, we describe how to reconstruct organ shapes using low-dimensional and high-dimensional latent variables. In Figure \ref{fig1} (b), EncBlock takes the mesh vertex group $\boldsymbol{x}$ as input and returns the mesh vertex group $\boldsymbol{\hat{x}}$, and the distribution of latent variables, including the mean $\boldsymbol{\mu}$ and variance $\boldsymbol{\sigma}^2$. The latent variable $\boldsymbol{z}$ in each layer is calculated as

\begin{equation}
    \boldsymbol{z} = \boldsymbol{\mu} + \boldsymbol{\epsilon} \boldsymbol{\sigma} 
    \label{eq: z}
\end{equation}

using $\boldsymbol{\epsilon}$, which is a random number sampled from a normal distribution, according to the reparameterization trick\cite{Kingma13}.
Note that $0$ is used for $\epsilon$ to ensure that identical outputs are obtained for identical inputs during inference. DecBlock takes the latent variable $\boldsymbol{z}$ received from EncBlock and outputs $\boldsymbol{x}$ received from DecBlock one level above as the input.

\subsubsection{Loss function}
To achieve highly accurate reconstruction while maintaining generalization performance, we introduce a loss function:
 \begin{equation}
\label{eq: whole_loss}
    \mathcal{L} = \mathcal{L}_{KL} + \alpha \mathcal{L}_{P},
\end{equation}

which combines the loss to the distribution of the latent variables and the input-output reconstruction error.

$\mathcal{L}_{KL}$ is the loss term for the distribution of the latent variable:

\begin{equation}
\label{eq: KLD}
    \mathcal{L}_{KL} = \mathcal{D}_{KL}( q(\boldsymbol{z}|\boldsymbol{x})|| \mathcal{N}(0,1)). 
\end{equation}

$\mathcal{D}_{KL}$ is the Kullback--Leibler divergence, where $\mathcal{N}(0,1)$ is the normal distribution and $q(\boldsymbol{z}|\boldsymbol{x})$ the distribution of the latent variable.

$\mathcal{L}_{P}$ is the mean squared error obtained from the corresponding vertex coordinates of the input and output vertex groups:
\begin{equation}
\label{eq: Pos}
    \mathcal{L}_{P} = \frac{1}{N}\frac{1}{M}\sum_{i=1}^{N}\sum_{j=1}^{M}||\boldsymbol{v}^{(i)}_j - \hat{\boldsymbol{v}_j}^{(i)}||_2^2.
\end{equation}

Let $N$ and $M$ be the number of vertices and cases in a single mesh, respectively, and $\boldsymbol{v}^{(i)}_j$ and ${\hat{\boldsymbol{v}_j}^{(i)}}$ be the 3D vectors representing the $i$-th vertex position of the $j$-th mesh of the correct and inferred data, respectively. $\mathcal{L}_{KL}$ makes the distribution of the latent variables as close as possible to the multivariate normal distribution and the mesh shape inferred by $\mathcal{L}_{P}$ as close as possible to the correct mesh shape. During training, the model weights are updated to minimize \eqref{eq: whole_loss}.

\section{Experiment}
To confirm the performance of the proposed hierarchical MeshVAE, experiments were conducted to quantitatively evaluate shape reconstruction performance for multiple organs in the abdomen and to confirm the shape interpolation representation, which is one of the functions of the organ shape atlas. Python 3.9, PyTorch, and PyTorch Geometric\cite{PyG} were used to implement the proposed model. The batch size in training was 32, the maximum number of epochs was set to 1000, and early stopping \cite{Morgan89} was used by monitoring loss $\mathcal{L}_{P}$ of the data for validation. The early stopping patience was set to 50 epochs and Adam\cite{Kingma14} with a learning rate of $1 \times 10 ^ {-2}$ was used to optimize the model. The performance of the model was checked for several combinations of hyperparameters of the loss function and $\alpha = 1.0 \times 10^{12} $ was used.

\subsection{Dataset and preprocess}
Abdominal 3D CT images of 124 patients who underwent radiotherapy for pancreatic cancer at the Department of Radiotherapy, Kyoto University Hospital, were the subjects of the experiment. First, surface meshes for each organ region were obtained from the contour information of the liver and stomach manually defined by a radiation oncologist. Hierarchical templates were then obtained by downsampling the number of vertices and subdividing the triangular mesh. Considering the size and complexity of the organ shape, three levels of 48, 168, and 738 vertices were used for the liver, and two levels of 168 and 738 vertices were used for the stomach because too large a reduction in the number of vertices during the generation of the low-resolution templates would result in the failure to represent organ features.

To obtain meshes with the same topological structure with vertex correspondence, mesh deformation registration\cite{Nakao21} was performed on the high-resolution meshes for all cases, which resulted in the group of registered liver and stomach meshes illustrated in Figure \ref{sample}. All mesh vertices were located in the range $[-256,256]$ (unit: mm) on the $x$-axis (left-right), $y$-axis (anterior-posterior), and $z$-axis (head-foot), with the origin at the position of the pancreatic tumor, that is, the target of the therapeutic beam in radiotherapy. The values of the coordinates were divided by 256 and normalized so that the range of the vertex coordinates was $[-1,1]$, which was used as the input for the model. A dataset of 124 subjects was split into 86 for training, 19 for validation, and 19 for testing. The datasets and experiments were approved by the Kyoto University Medical Ethics Committee (approval number: R1446).

\subsection{Evaluation value}

The following three indicators were used for the quantitative evaluation of the results:

\begin{itemize}
    \item mean absolute error (MAE)
    \item mean distance (MD)
    \item Hausdorff distance (HD)
\end{itemize}

The MAE is calculated to quantify the error between the vertices of the mesh based on

\begin{equation}
    E_{MAE} = \frac{1}{NM}\sum_i^M\sum_j^N||\boldsymbol{v^{(j)}_i}-\boldsymbol{v'^{(j)}_i}||_2
    \label{eq: MAE}
\end{equation}

where $N$ denotes the number of mesh vertices, $M$ denotes the number of cases to be evaluated, and $\boldsymbol{v}$ and, $\boldsymbol{v}'$ denote the vertex positions of the mesh to be restored and predicted, respectively.  MD\cite{Riguad18}\cite{Kim14} and HD\cite{Heimann092} are indices that quantify the misalignment between the surfaces of the two meshes. HD is the distance between the two meshes where the mean shape deviates the most between the two meshes, whereas, MD quantifies the average deviation between the surfaces of the two meshes.

\subsection{Mesh reconstruction performance comparison}

First, experiments were conducted to check the mesh reconstruction performance of several models and the proposed method. Table \ref{tab} show the reconstruction performance of each model for the liver and stomach shapes, respectively. In the table, FC represents a model with only fully connected layers, GCN represents a model constructed using only the ResBlocks in Figure\ref{fig1}, and Pooling represents a model that combines ResBlocks and SubdivPooling/Unpooling, and compresses the dimensions of the latent variables. The MAE, MD, and HD indices were calculated for training, validation, and test data.

\begin{table*}[t]

\renewcommand{\arraystretch}{1.5}
\centering
\caption{Comparison of liver and stomach mesh reconstruction performance for each model. Mean of MAE, MD and HD for train data (Train), validation data (Val) and test data (Test)}
    \begin{subtable}{
        \begin{tabular}{*{10}{c}}
        \hline
        \begin{tabular}[c|]{@{}c@{}}Liver \end{tabular} &
        \multicolumn{3}{c}{\begin{tabular}[c]{@{}c@{}}MAE(mm) \end{tabular}} & 
        \multicolumn{3}{c}{\begin{tabular}[c]{@{}c@{}}MD(mm) \end{tabular}} & 
        \multicolumn{3}{c}{\begin{tabular}[c]{@{}c@{}}HD(mm) \end{tabular}}\\ 
        \cmidrule(r){2-4} \cmidrule(r){5-7} \cmidrule(r){8-10}
        Model & Train & Val & Test & Train & Val & Test & Train & Val & Test\\
        \hline
        PCA & 25.26 & 25.67 & 23.46 & 12.32 & 12.80 & 11.36 & 35.02 & 36.05 & 37.18\\
    
        FC & 7.69 & 14.52 & 13.60 & 3.77 & 7.19 & 6.38 & 14.71 & 26.11 & 25.38\\
    
        GCN & 14.82 & 14.38 & 14.48 & 8.81 & 8.29 & 8.43 & 34.15 & 36.39 & 34.16\\
    
        Pooling & 2.91 & 2.93 & 2.95 & 1.38 & 1.39 & 1.40 & 8.86 & 8.66 & 8.36\\
    
        \textbf{Proposed} & \textbf{1.49} & \textbf{1.52} & \textbf{1.50} & \textbf{0.76} & \textbf{0.79} & \textbf{0.74} & \textbf{7.09} & \textbf{7.58} & \textbf{6.78}\\
        \hline
        \end{tabular}
        }
    \end{subtable}
    
    \begin{subtable}{
        \renewcommand{\arraystretch}{1.5}  
        \centering
        \begin{tabular}{*{10}{c}}
        \hline
        \begin{tabular}[c|]{@{}c@{}}Stomach \end{tabular} &
        \multicolumn{3}{c}{\begin{tabular}[c]{@{}c@{}}MAE(mm) \end{tabular}} & 
        \multicolumn{3}{c}{\begin{tabular}[c]{@{}c@{}}MD(mm) \end{tabular}} & 
        \multicolumn{3}{c}{\begin{tabular}[c]{@{}c@{}}HD(mm) \end{tabular}}\\ 
        \cmidrule(r){2-4} \cmidrule(r){5-7} \cmidrule(r){8-10}
        Model & Train & Val & Test & Train & Val & Test & Train & Val & Test\\
        \hline
        PCA & 26.58 & 32.02 & 24.31 & 13.07 & 15.48 & 11.83 & 36.02 & 42.71 & 33.68\\
    
        FC & 6.15 & 14.89 & 12.82 & 2.96 & 7.14 & 5.94 & 11.48 & 25.10 & 22.64\\
    
        GCN & 14.89 & 14.65 & 15.13 & 11.08 & 11.02 & 11.32 & 40.02 & 41.75 & 40.46\\
    
        Pooling & 2.72 & 2.93 & 2.61 & 1.31 & 1.44 & 1.24 & 6.57 & 7.45 & 6.33\\
    
        \textbf{Proposed} & \textbf{1.31} & \textbf{1.46} & \textbf{1.44} & \textbf{0.68} & \textbf{0.76} & \textbf{0.81} & \textbf{5.36} & \textbf{6.83} & \textbf{6.57}\\
        \hline
        \end{tabular}
        }
    \end{subtable}

\label{tab}
\end{table*}

In Table \ref{tab}, PCA had poor reconstruction performance, and the other deep learning-based methods achieved good accuracy on all indices and data. For FC, the validation and test indices significantly outperformed the training indices. This may be because of overfitting during training caused by the large number of parameters in the fully connected layer. By contrast, no significant performance differences were observed for GCN, Pooling, and Proposed for training, validation, and test data, which suggests that overfitting did not occur for these models. A comparison of GCN and Pooling showed a significant improvement in Pooling, which confirmed that the proposed SubdivPooling/Unpooling contributed to shape reconstruction. Compared Proposed to Pooling also confirmed that the hierarchization of latent variables was effective in improving representation performance.

A comparison of Table \ref{tab} shows that the performance of each model for the liver and stomach shapes did not differ significantly, even though the stomach shape was smaller than the liver shape. This may be because the stomach has a more complex shape than the liver, and the shape differences between cases are large and difficult to learn.

\begin{figure}[t]
	\centering
	\includegraphics[width=1.0\linewidth]{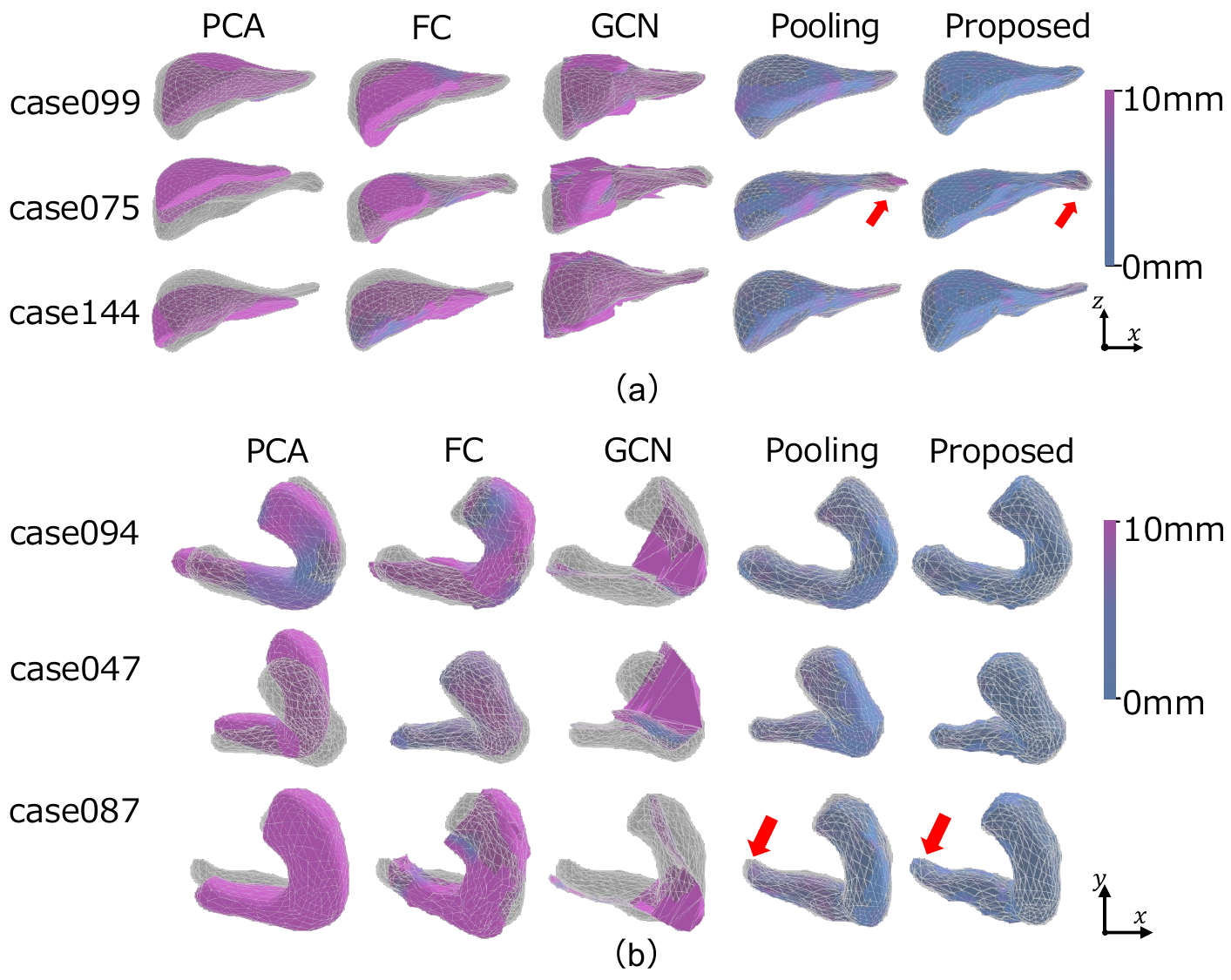}
	\caption{Result of the reconstruction of the organ shape for each model: (a) liver, (b) stomach. The gray mesh represents the correct data and the colored mesh represents the prediction result. The difference between the correct and predicted position of each vertex is colored.}
	\label{liver-samples}
\end{figure}

Figure\ref{liver-samples} shows three examples of the reconstruction results for the liver and stomach shapes for each model. The gray mesh in the figure represents the correct shape to be reconstructed and the colored surface mesh is the result of the reconstruction by each model. For each vertex, the error is calculated as the distance between the correct and inferred positions, and the surface is colored according to the error using the color map shown in the figure. The cases shown in the figure were selected from the first, second, and third quartiles from the left in the reconstruction of organ shapes in the test data using the proposed method, in ascending order of MD value. In Figure\ref{liver-samples}, the center of gravity of each mesh was aligned and the sizes were normalized to help a comparison of the shapes.

For Figure\ref{liver-samples} (a), the prediction results in PCA were close to the template, with a large positional error. FC recovered a shape close to the correct data. However, the mesh position error was large. GCN, Pooling, and Proposed reconstructed the shape with fewer positional errors than FC. However, GCN obtained many outputs with sharp surfaces and local features were not well represented. Pooling and Proposed, which introduced SubdivPooling, reconstructed organ shapes more accurately than the other methods. As shown by the red arrows in the figure, Proposed reconstructed the shape more accurately than Pooling, even in areas with large curvature. This can be considered as the result of learning to separate local and global features using hierarchical latent variables.

Figure \ref{liver-samples} (b) shows the results of stomach shape reconstruction. The prediction results for PCA and FC demonstrated similar trends to those for liver shape reconstruction. The output of GCN was a mesh with a collapsed shape, although the position of the center of gravity was reproduced, which suggests that the reconstruction of the stomach shape was more difficult than that of the liver shape. Pooling and Proposed reconstructed the shape of the organ more accurately than the other methods. A comparison of Pooling and Proposed showed that Proposed reconstructed the shape better in areas of large curvature, in addition to the shape of the liver. Additionally, the position error of the reconstruction result of Pooling was larger than that of Proposed. There were no significant differences in performance between the cases for the liver and stomach shapes.

\begin{figure*}[htbp]
	\centering
	\includegraphics[width=1.0\linewidth]{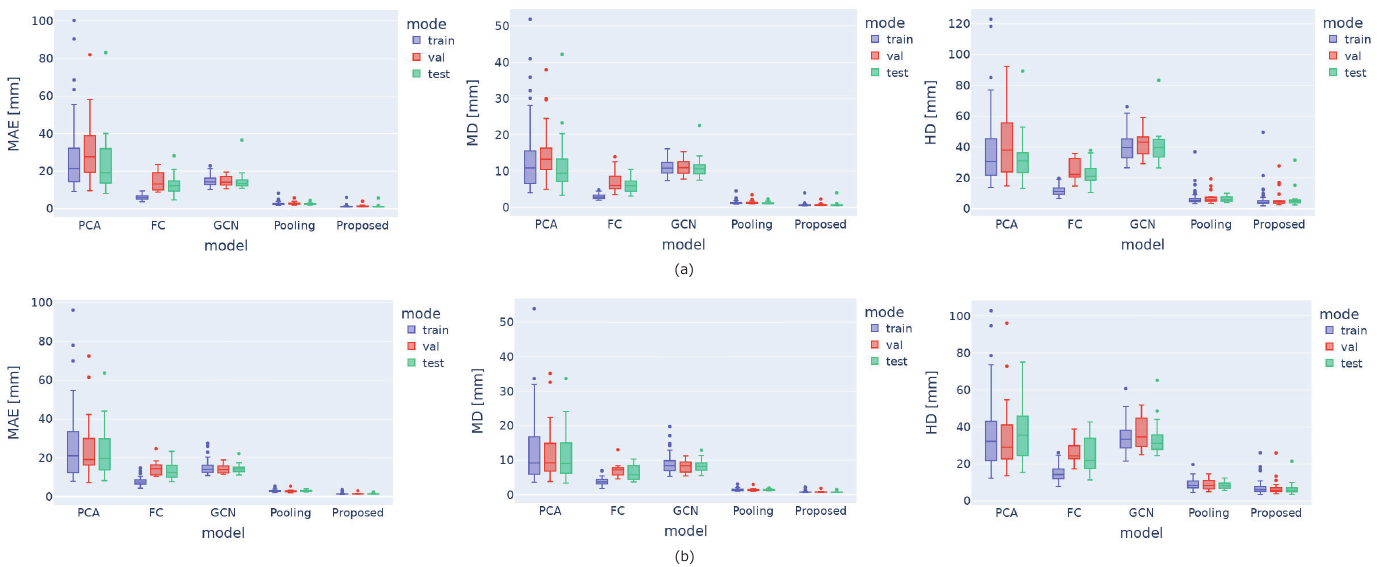}
	\caption{Performance of mesh reconstruction: (a) liver MAE,  MD, HD, (b) stomach MAE, MD, HD.}
	\label{liver-boxplot}
\end{figure*}

Figure\ref{liver-boxplot} shows box-and-whisker plots of the reconstruction performance of each model for liver and stomach data. Figure (a) shows the results of liver for MAE, MD, and HD, respectively, with the training, validation, and test datasets shown in blue, red, and green respectively. The prediction results for PCA showed larger quartile deviations of the indices for all datasets compared with the other methods. Pooling, Proposed were able to reconstruct organ shapes with higher accuracy compared with the other methods. From the box-and-whisker diagram of the liver shape prediction results shown in Figure\ref{liver-boxplot} (a), Proposed did not sufficiently recover some organ shapes and vertex positions because there were more cases with large values in HD than in MAE. From the box-and-whisker diagram of the prediction results for the stomach shape shown in Figure\ref{liver-boxplot} (b), there were cases where all methods had larger errors than in the case of the liver shape. This is believed to be because the stomach had large shape differences between cases, and as shown in \eqref{eq: z}, VAE reconstructed the shape based on the mean $\mu$ in the sampling of latent variables, which tended to output the mean shape.

\subsection{Hierarchical latent variable shape interpolation representation.}

Finally, an experiment was conducted to check the shape interpolation representation by changing the hierarchical latent variables as a function of the organ shape atlas achieved by the proposed model. The objectives of this experiment were threefold.

\begin{itemize}
    \item to confirm the organ shape generated by interpolating the latent variables obtained from different cases
    \item to confirm the change in organ shape by changing different hierarchical latent variables
    \item to compare the shape interpolation representation by PCA and the proposed method.
\end{itemize}

In the shape interpolation representation in PCA, the eigenvalues were arranged in order of contribution, with the single component with the largest contribution and the other nine components as the low-resolution component $z_2$ and high-resolution components $z_1$. In Proposed, the low-resolution and high-resolution latent variables obtained by inputting the mesh to the encoder are $z_2$ and $z_1$, respectively.

After the latent variables $z_1^{(A)}, z_2^{(A)}$ and $z_1^{(B)}, z_2^{(B)}$ were obtained from the two cases $A$ and $B$, in (\ref{eq: morphing}), $\alpha,\beta$ were updated at regular intervals from 0 to 1 and the latent variables were continuously changed from $z_1^{(t)}$ to $z_2^{(t)}$ in latent space. The obtained $z_1^{(t)}, z_2^{(t)}$ were used to reconstruct 3D shape.

\begin{equation}
\left\{
\begin{array}{l}
    z_1^{(t)} = (1 - \alpha)z_1^{(A)} + \alpha z_1^{(B)} \\
    z_2^{(t)} = (1 - \beta)z_2^{(B)} + \beta z_2^{(B)}
\end{array}
\right.
    \label{eq: morphing}
\end{equation}

\begin{figure}[t]
	\centering
	\includegraphics[width=1\linewidth]{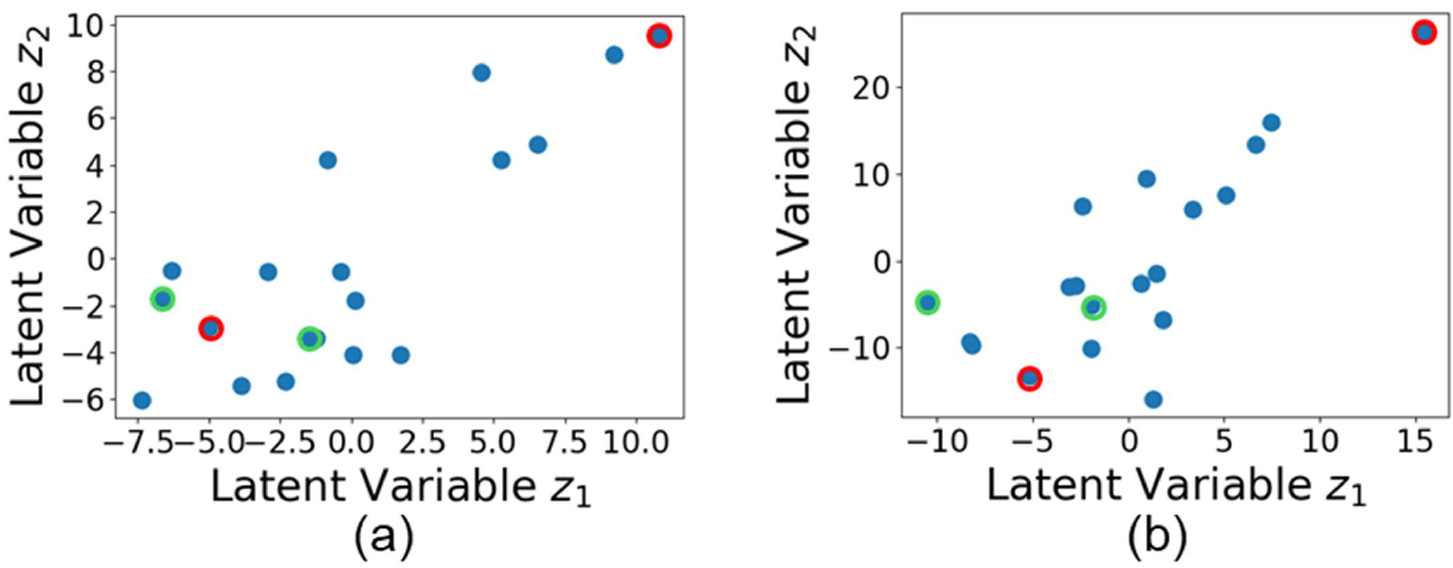}
	\caption{Scatter plots of the hierarchical latent variables of the organ shape mesh obtained by training the proposed model: (a) liver and (b) stomach. The corresponding circles indicate the two cases subjected to shape interpolation.}
	\label{latent-scatter}
\end{figure}

The latent variables $z_1, z_2$ of the stomach shape test data sampled by the proposed method were each dimensionally compressed to one dimension by PCA and illustrated as scatter plots in Figure \ref{latent-scatter}.

In the experiment, the following two sets of cases were selected, whose shapes were very different and whose positions were far apart in the scatter plots in Figure\ref{latent-scatter}. Liver and stomach shapes, respectively, were represented by the proposed method and shape interpolation using PCA:
\begin{itemize}
    \item red circles indicate case 055 and case 090
    \item green circles indicate case 049 and case 132.
\end{itemize}

\begin{figure*}[t]
	\centering
	\includegraphics[width=0.88\linewidth, trim=0cm 0cm 0cm 0cm, clip]{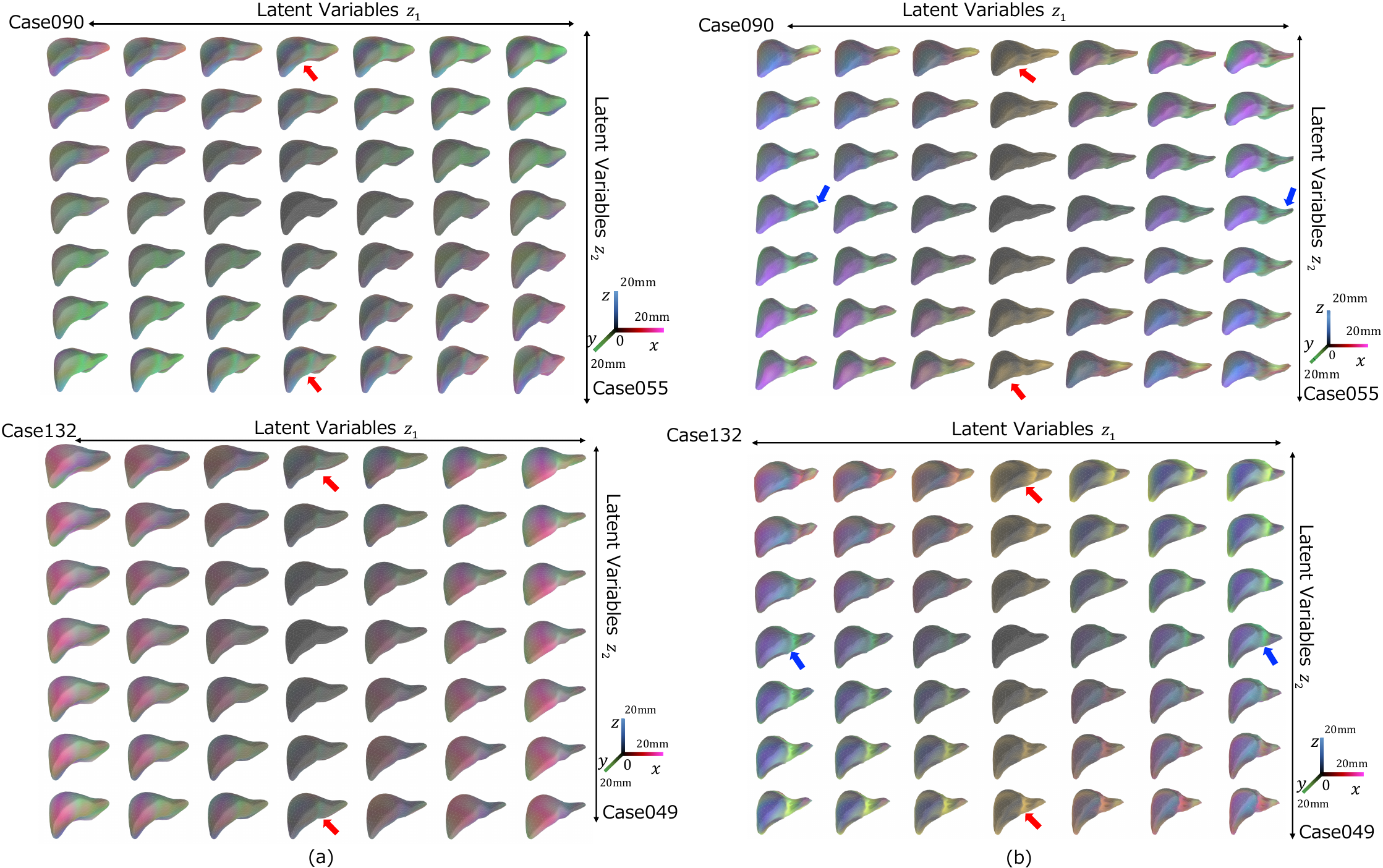}
	\caption{Shape interpolation of the liver using PCA and the proposed method. (a) PCA, (b) proposed method, showing the shape interpolation of two groups: case090-case055 and case132-case049.}
	\label{liv-morphing1}
\end{figure*}

\begin{figure*}[b]
	\centering
	\includegraphics[width=0.88\linewidth]{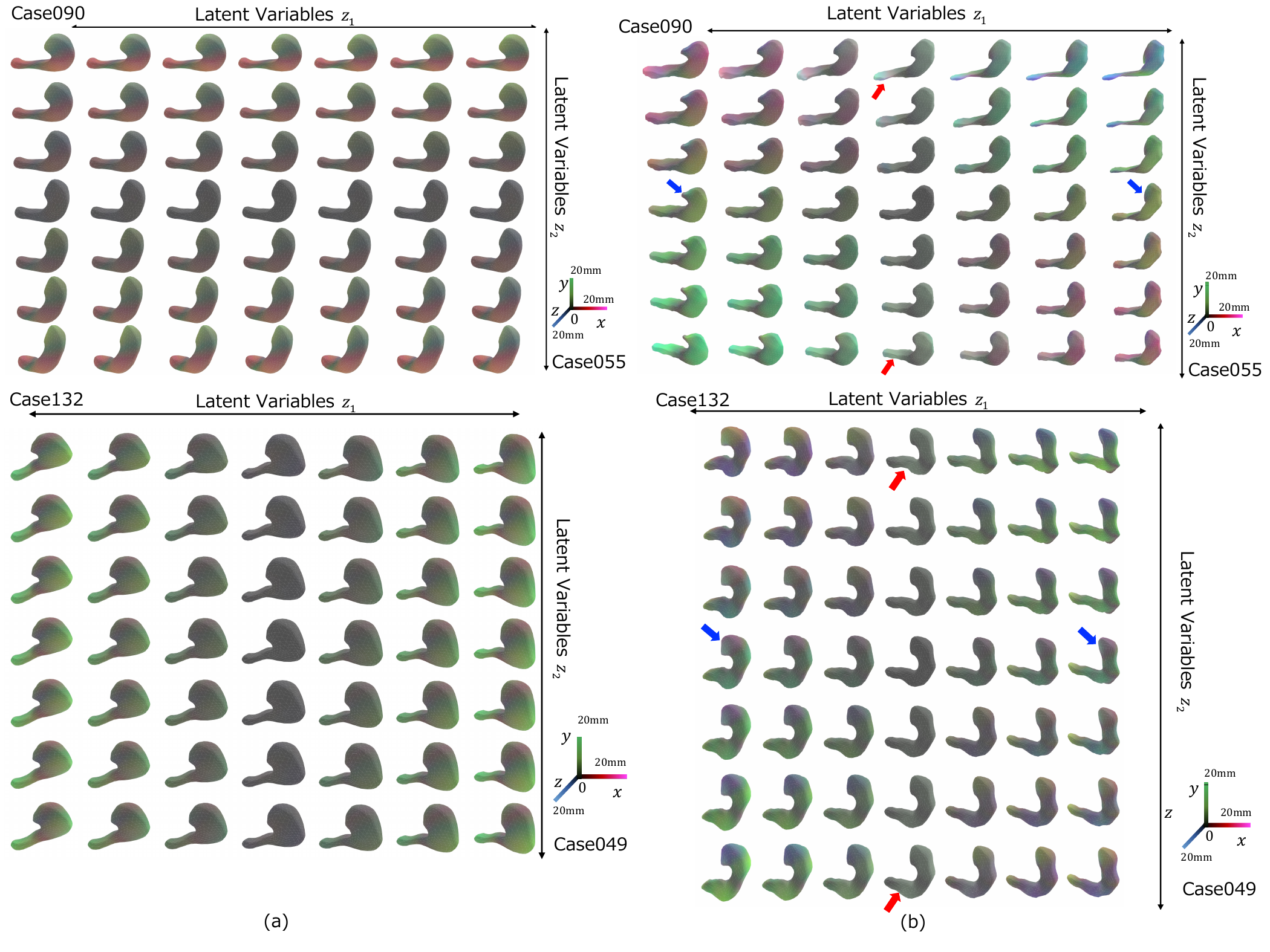}
	\caption{Shape interpolation of the stomach using PCA and the proposed method. (a) PCA, (b) proposed method, showing the shape interpolation of two groups: case090-case055 and case132-case049.}
	\label{st-morphing1}
\end{figure*}

Figures \ref{liv-morphing1}, and \ref{st-morphing1} show the shape interpolation results between the two selected cases of liver and stomach shapes. (a) and (b) in each figure show the shape interpolation representation by PCA and the proposed method, respectively. The gray mesh in the middle of the diagram is the intermediate shape in the latent space of the two cases. The shapes obtained by changing the latent variables $z_1$ and $z_2$ were colored according to the RGB values to make the displacements in each axis direction from the intermediate shape easily visible. Each mesh was normalized so that the center of gravity was in the middle, and the displacements were calculated and visualized by focusing only on the shape change.

First, (b) in Figures \ref{liv-morphing1} and \ref{st-morphing1}, confirmed that the proposed method generated an interpolated shape between two cases by changing the latent variable and represented smooth shape changes.

Focusing on the area indicated by the red arrow in the shape interpolation representation of the liver in Figures \ref{liv-morphing1}(a) and \ref{liv-morphing1}(b) indicates that the proposed method represented nonlinear interpolation because the entire shape changed with rotation, whereas PCA represented a linear transformation, similar to scaling in the $x$-axis direction. Additionally, the shape of the liver tip was significantly altered by the proposed method and local shape changes were represented by high-resolution latent variables. Near the red arrow in Figure \ref{liv-morphing1}(b), a global change occurred that was similar to scaling in the $x$-axis direction, whereas near the blue arrow, the curvature of the shape surface changed. In this example, the proposed method achieved shape interpolation that separated the global and local shape features.

Figure \ref{st-morphing1} shows the results of shape interpolation between the two cases for the stomach shape. The shape of the stomach is smaller than that of the liver, and there is a large difference in shape between patients because of differences in body shape and the presence or absence of contents. The interpolated representation by PCA in (a) shows that the entire shape changed when the low-resolution component was updated, whereas the shape barely changed when the high-resolution component was updated. By contrast, in the interpolated representation using the proposed model in (b), when the low-resolution latent variable $z_2$ was updated, the curvature of the organ changed, as shown by the red arrow. When the high-resolution latent variable $z_1$ was updated, the thickness of the organ changed significantly, as shown by the blue arrow. These results suggest that local features of the organ shape were encoded in the high-resolution latent variables and global features of the organ shape were encoded in the low-resolution latent variables through training the model. Compared with PCA, the proposed method represented shape interpolation with hierarchically separated shape features and the effect of nonlinear shape interpolation in latent space was more pronounced for the shape of the stomach than that of the liver.

\section{Conclusion}
In this study, we introduced hierarchical latent variables into a mesh variational autoencoder with the aim to construct an organ shape atlas with both high reconstruction performance and interpretability. For hierarchization, we proposed a hierarchical template, a SubdivPooling/Unpooling operation based on mesh subdivision, and hierarchical latent variables. The proposed method enables the hierarchical analysis of latent variables of shape differences between patients by hierarchizing latent variables, which improves the interpretability of nonlinear organ shape representations. To confirm the effectiveness of the proposed method, we validated the model constructed on organ mesh data of the liver and stomach from 124 cases. For the test data of 19 of these cases, the mean distance between vertices was 1.5 mm, the mean distance for the liver shape was 0.7 mm, the mean distance between vertices was 1.4 mm, and the mean distance for the stomach was 0.8 mm. We performed two sets of shape interpolation between two cases for the liver shape and stomach shape, and compared the shape interpolation expressions for the proposed method and the conventional method using PCA. We confirmed that the proposed method represented the interpolated shape continuously in the shape interpolation representation and that by changing the latent variables in different layers, the proposed method represented the shape interpolation by hierarchically separating the shape features unlike PCA.

Future work will include a deeper hierarchy, a structure that can control the output with fewer latent variables, and a framework that enables shape interpolation with a missing part of the mesh for clinical purposes.

\section*{Acknowledgment }
We thank Edanz (https://jp.edanz.com/ac) for editing a draft of this manuscript.

\end{document}